\documentstyle[preprint,prl,aps]{revtex}

\def\beq{\begin{equation}}
\def\eeq{\end{equation}}
\def\beqa{\begin{eqnarray}}
\def\eeqa{\end{eqnarray}}
\def\lla{\left\langle}
\def\rra{\right\rangle}

\def\lsim{\mathrel{\raise.3ex\hbox{$<$\kern-.75em\lower1ex\hbox{$\sim$}}} }
\def\gsim{\mathrel{\raise.3ex\hbox{$>$\kern-.75em\lower1ex\hbox{$\sim$}}} }

\begin{document}
\addtolength{\baselineskip}{-.3\baselineskip}
\draft
\preprint{{\vbox{\hbox {UR-1571} \hbox{Nov 1999}\hbox{rev. Jan 2000}}}}
%@twocolumnfalse\endcsname

\title{Pseudo-Dirac Neutrinos  }
\author{\bf Darwin Chang$^{1}$
$\!$\footnote{E-mails: chang@phys.nthu.edu.tw}
and Otto C. W. Kong$^2$
$\!$\footnote{E-mails: kongcw@phys.sinica.edu.tw}
}
\address{$^1$ NCTS and Physics Department, N. Tsing-Hua University,\\
Hsinchu, Taiwan, R.O.C. \\
$^2$Department of Physics and Astronomy,\\
University of Rochester, Rochester NY 14627-0171,\\
and \\Institute of Physics, Academia Sinica, Nankang, Taipei, R.O.C.$\!$
\footnote{Present address}.}
%\date{\today}
\maketitle

%\vskip -0.5cm
\begin{abstract}
We propose a scheme in which a pseudo-Dirac structure for three family of
light neutrinos is generated naturally. An extended Higgs sector
with a majoron is used for the generation of the leptonic number
violating neutrino Majorana mass. The resultant neutrino mass matrix could
easily fit all available experimental data. We discuss relevant
constraints on the scales involved for the model to be phenomenologically
viable.
\end{abstract}
\pacs{}
%\vskip0pc]
%\vskip2pc]

%\newpage
%\narrowtext

{\it Introduction.}
It is well known that the fermion spectrum of the Standard Model (SM)
exhibits a three-family structure with a strong hierarchy of masses among
them.  In the quark sector, this has been taken as an indication of
hierarchical mass matrices with small mixing among the families, as for
example discussed in Ref.\cite{RRR}. The charged lepton masses show
the same hierarchy, while no physical mixing parameters can be introduced
without first giving masses to neutrinos, which cannot happen within the
SM.

Recently, experimental data on atmospheric and solar neutrinos strongly
suggest the existence of neutrino masses and
oscillations\cite{BKS,sK,LSND}.  Furthermore, taking seriously all the available
experimental data altogether suggests the existence of more than three
light neutrinos\cite{4nu}.  As indicated by the atmospheric neutrino oscillation
data\cite{sK}, if there are only three active neutrinos, as in the SM,
a different kind of family mixing or flavor structure would be
implied\cite{EIR,kong,gut}.  In particular, the mixing angle that
is responsible for the atmospheric neutrino oscillation is determined to
be maximal.  It is also very likely that the mixing angle that is
responsible for the solar neutrino deficit is also nearly maximal, be it
large angle MSW resonant oscillation solution or vacuum
oscillation. This naturally
poses a puzzle : why are neutrino family hierarchy so different from that of
the quarks or the charged leptons?  Indeed, there are many papers in the
literature that address this disparity in the hierarchical structures.
Some works within three active family framework and use intricate lepton
family symmetries or quantum correction to generate the new hierarchy in
the neutrino sector.  Alternatively, one can introduce the right-handed
neutrinos (in parallel with the quark and charged lepton sectors) and
use the fact that lepton number can be broken uniquely in the neutrino
sector to blame the new hierarchy on the related Majorana masses of the
right-handed neutrinos.  In that case the Dirac masses may
still enjoy the same hierarchy as the quark and charged lepton
sectors.  This actually may be a more practical approach toward the
various problems facing particle physics.  It is commonly recognized that
family hierarchy problem is very tough to elucidate.  This approach
allows one to decouple the family hierarchy problem with some of the
problems related to neutrino oscillations such as the smallness of neutrino
masses and the maximal character of some of the mixing angles.
Furthermore, another hint that support this picture of neutrino mass
pattern is the the data from LSND\cite{LSND} which suggests that the muon
neutrino mixes with some other neutrino with a mixing angle  the size of
the cabibbo angle.  This would follow from the current picture if the Dirac
masses of the neutrino have the same hierarchical structure as that of
the quarks and charged leptons.

Three families of right-handed neutrinos
is a common feature of many extensions of
SM, in an $SO(10)$ unification framework or otherwise\cite{3nuR}.
Among the extension of SM with right-handed neutrinos, the
most popular ones incorperate the so-called ``seesaw" structure, which
naturally explain the smallness of the neutrino masses by invoking a high
intermediate or GUTs scale.  However, unless the issue is coupled with the
fermion hierarchy problem, this approach does not naturally explain why
some of the oscillation angles are maximal. Nor can it provide a
fourth light neutrino.   In this letter, we take a
different approach.  We assume that both the atmospheric and the solar
neutrino oscillation angles are nearly maximal and investigate how this
can achieved in a model with {\it light} right-handed neutrinos,
or better known as sterile neutrinos.
We discuss a scenario in which the resulting neutrino masses is naturally
pseudo-Dirac and compatible with the usual hierarchical mass
structure of quarks and charged lepton sectors.
We will show that this scenario could be a natural consequence of a
slightly broken lepton number symmetry, and naturally explains the
experimental data.   In addition to giving rise to the maximal mixing
angles in atmospheric and solar neutrino mixings naturally, it also
naturally explains by the $\nu_e$-$\nu_\mu$ mixing angle observed in the
LSND experiment is close to the Cabbibo angle.

The greatest potential problem about having more sterile neutrinos lies with
big bang nucleosynthesis (BBN)\cite{BBN}, as the latter predictions are quite
sensitive to the $N_\nu$, the effective number of neutrino species.
However, as suggested in Ref.\cite{FV}, the BBN bounds on mixing of
neutrinos with sterile species can be considerably weakened in the
presence of relatively large neutrino asymmetry. Furthermore, it has also
been shown that active-sterile oscillation in one channel could
be used to create such an asymmetry which then suppresses the corresponding
mixing in other channel(s). Detailed studies \cite{xSF} has established
the feasiblity of the scenario, with only relative weak constraint on the
neutrino parameter\footnote{
Perhaps it should be mentioned that calculation from one other group\cite{SF}
arrived at a substantially different result, putting a much stronger
constraint on the scenario.}.
Such studies typically concentrates on
small neutrino masses.  A complementary
studies of the heavy  $\nu_{\scriptscriptstyle \tau}$ scenario\cite{DHPS}
shows that the contribution of the latter to $N_\nu$ could go as far as
$-2$ for some value of decay lifetime. Meanwhile, other authors\cite{KS}
suggest that  $N_\nu$ can be as high as $4.53$. All these certainly suggests
a generic situation with three sterile neutrinos is not definitely ruled out,
while  more carefully analyses along the line of Ref.\cite{xSF} have to be performed to check the BBN constraint on a specific model.

Another major problem of the pseudo-Dirac idea is the indications from
recent statistical analyses of all available solar neutrino data\cite{BKS}
that the large mixing angle solution into a sterile neutrino is ruled out
as a solution. However, as argued in Ref.\cite{CFV}, the Chlorine experiment
result is actually in general disagreement with other experiments, and if
the former is left out, maximal
$\nu_{\scriptscriptstyle e} \longrightarrow \nu_{\scriptscriptstyle s}$
solution to the solar problem is constrained but still interesting.
Hence, we consider the pseudo-Dirac idea still worths some attention.
In fact, after the completion of our model, we realized that there has been
quite some discussion in the literature about the general scenario\cite{pD},
as well as some model building works. The latter includes the mirror
matter model\cite{mir} and more from some other more general
settings\cite{pDm}. Our model has the special feature that it uses the
(approximate) lepton number symmetry itself to enforce the pseudo-Dirac
mass pattern and a Dirac seesaw structure for the suppressions of the
dominating Dirac masses from the electroweak scale. Hence, we consider it
an interesting alternative.

{\it  Pseudo-Dirac mass scheme.}
The generic Majorana mass matrix with three
$SU(3)\times SU(2)\times U(1)$ singlet fermions, {\it i.e.} neutrinos,
is given by
\begin{equation} \label{mass}
\left(\begin{array}{cc}
M_\nu & D \\
D^T & M
\end{array}\right)\; ,
\end{equation}
where $M_\nu$ and $M$ are the lepton number violating Majorana
masses of the $\nu_{\scriptscriptstyle L}$'s  and
$\nu_{\scriptscriptstyle S}$'s respectively, and $D$ their Dirac masses.
The popular seesaw scenario is characterized by $M_\nu \ll D \ll M$.
However, here we shall adopt a different scenario and look for scheme that
can naturally generate a mass matrix which is of pseudo-Dirac form by
having $M_\nu\; , M \ll D$.

The philosophy here is to generate all the small quantities through some
kind of seesaw mechanism associated with a larger scales.  The smallness
of $M_\nu$ and $M$ can be taken as a consequence of an approximate lepton
number symmetry.  When lepton number is spontaneously broken, the
smallness of the associated vacuum expectation value (VEV) can be arranged
naturally through the scalar seesaw mechanism\cite{MS}, as we shall
demonstrate below.  In order to explain the smallness of $D$, we will have
to invent new mechanism.  One such mechanism called Dirac seesaw mechanism
will be illustrated below.

Before we discuss the mechanism, let us first discuss how the
pseudo-Dirac mass scheme fits the experimental data, and what its
predictions would be. The basic idea here is motivated by the experimental
indications of small mass-squared differences and near maximal mixings for
$\nu_{\scriptscriptstyle \mu} \longrightarrow \nu_{\scriptscriptstyle X}$
and probably also for $\nu_{\scriptscriptstyle \mu} \longrightarrow
\nu_{\scriptscriptstyle e}$, with a relatively ``larger" mass-squared
difference.  The latter feature suggests resemblance with the familiar SM
flavor structure. A pure Dirac
mass contribution, as would be enforced by lepton number conservation,
gives pairs of Majorana neutrino mass eigenstates of equal and opposite
masses and $45^o$ mixings. If $D$, $M_\nu$, and  $M$ were all nearly diagonal, in
analogy with the quark and charged lepton mass matrices, one should be
able to account for atmospheric neutrino oscillation as $\nu_\mu$ into its
sterile partner and account for  solar
neutrino oscillation as $\nu_e$ into its sterile partner.
In the latter case, close to maximal mixing is still preferred. However,
dropping the Chlorine experiment result also implies an extended range of
acceptable parameters beyond the standard large angle MSW or vacuum
oscillation solutions\cite{CFV}. The respective
small mass differences are explained by small $M_\nu$ and $M$.
Small Majorana mass contributions, will lift the pairwise mass degeneracy
(neglecting the unphysical sign difference) and shift
the mixing angles away from $45^o$. Hence, each
$\nu_{\scriptscriptstyle L}$ will have
a nearly degenerate and maximally mixed sterile partner, while the full
neutrino spectrum still follows a hierarchical pattern similar to the other
SM fermion: $m_{\nu_{\scriptscriptstyle e}} \ll
m_{\nu_{\scriptscriptstyle \mu}} \ll m_{\nu_{\scriptscriptstyle \tau}}$
with small family mixings.
In addition, the Cabbibo size mixing observed by LSND will be explained by
the flavor mixing in analogy with the regular quark sector.
This is the basic picture which we consider very natural and straight
forward.

Before we propose the generating mechanism for the mass matrix let
us see what kind of values one needs to achieve for a satisfactory
model. The LSND data prefers $\Delta m^2 =
0.2\;\mbox{---}\;20\,\mbox{eV}^2$ and $\sin^2\!2\theta
=0.03\;\mbox{---}\;0.001$. The Super-K atmospheric neutrino data
prefers $\Delta m^2 = (0.5\;\mbox{---}\;6)\times
10^{-3}\,\mbox{eV}^2$ and $\sin^2\!2\theta >
0.82\;(90\%\,\mbox{C.L.})$. The solar neutrino data prefers two
large angle solutions: the MSW solution with $\Delta m^2 \approx
1.8 \times 10^{-5}\,\mbox{eV}^2$ and $\sin^2\!2\theta
\approx0.76$, and the vacuum oscillation solution with $\Delta m^2
\approx 6.5 \times 10^{-11}\mbox{eV}^2$ and $\sin^2\!2\theta
\approx 0.75$, which may have problem when a sterile neutrino is
involved. When the Chlorine experiment result is dropped, however,
analysis in Ref.\cite{CFV} finds an acceptable $\Delta m^2$ range
of $10^{-3}$-$10^{-10}\,\mbox{eV}^2$  with the interval $8\times
10^{-6}$-$2\times10^{-7}\,\mbox{eV}^2$ ruled out by day-night
asymmetry data. Taking these into account, a satisfactory spectrum
of of neutrino masses may be given as follows :
\begin{center}
\begin{tabbing}
aaaa \= aaaa \= aaaaaaaaaaaaa \=  $\delta\!m$ (eV)aaaaaa
\= aaaaade \kill
\>  \>   $m_{\nu}\;(\mbox{eV})$
 \>  $\Delta m_{\nu}$ (eV) \> $\Delta m_{\nu}^2 = 2 m_{\nu} \Delta
m_{\nu} \;
(\mbox{eV}^2)$ \\
\vspace*{.1in}
\> $ \nu_{\scriptscriptstyle e}$ \> $\sim  10^{\mbox{\rm -}2} \;\mbox{eV}$
 \> $\sim 10^{\mbox{\rm -}8}  $
\> $\sim 10^{\mbox{\rm -}10}  $\\
\> $ \nu_{\scriptscriptstyle \mu}$ \> $\sim 1\; $
 \> $\sim  10^{\mbox{\rm -}3} $
\> $\sim  10^{\mbox{\rm -}3} $  \\
\> $ \nu_{\scriptscriptstyle \tau}$ \> $\sim 20 $
 \> $\sim  10^{\mbox{\rm -}1} $
\>  $\sim 1$
\end{tabbing}
\end{center}
Here, in the $m_{\nu}$ column, the value for $m_{\nu_\mu}$ is motivated by
LSND result.
The other entries for $m_{\nu}$ are obtained by naively assuming that mass
hierarchy among the Dirac masses are similar to that of the charged leptons.
They should be considered only as suggestive values.
In the $\Delta m_{\nu}^2$ column, the rough value for
$\Delta m_{\nu_\mu}^2$ is given by recent atmospheric neutrino oscillation
data.  The rough values for $\Delta m_{\nu_e}^2$ is given by recent solar
neutrino data.  Here we take a value in the vacuum oscillation solution
range just for the illustration purpose.  Note that a large range of $\Delta
m^2_{\nu_e}$ is possible. As for $\nu_\tau$,
there is no direct experimental hint on $\Delta m_{\nu_\tau}$.
For instance, even $m_{\nu_{\scriptscriptstyle \tau}}$  in the MeV range
is a possibility. Much smaller value of
$m_{\nu_{\scriptscriptstyle \tau}}$ seems to fit more easily into a
specific pattern with the rest of the neutrino spectrum.
We choose  in the above
ratios among the Majorana masses (equivalent to the mass splittings,
$\Delta m$) to be roughly the squares of that of
among the charged leptons. This mass ratio pattern is
inspired by a simple approximate flavor symmetry perspective\cite{afs}
taken here only with the SM fermions. Note that the pattern
suggests a $\Delta m^2$ of the order
$1\,\mbox{eV}$ between $\nu_{\scriptscriptstyle \tau}$ and its sterile partner.

The above neutrino mass spectrum is presented only to give typical values
in order to provide an  order of magnitude estimate of various scales
that will be introduced later in our mechanism.
Our model will not, however, address the origin of the
hierarchy down the families, but only illustrate a mechanism for achieving
the pseudo-Dirac mass scheme.

{\it Scalar potential and Higgses.}
To have an approximate lepton number symmetry in the neutrino
sector, we introduce additional Higgs scalar(s) and demand that the lepton
number be broken only spontaneously or softly.
We wish to build a scheme such that the smallness of $M_\nu$ and $M$ are
due to the smallness of the corresponding lepton number breaking VEV's.
The smallness of these VEV's will be in turn explained naturally by the
existence of a larger scale.  While we can not explain the origin of
such scale, the tie between the small VEV and the larger scale is
natural.  If the lepton number symmetry is broken spontaneously, there will be
Majoron. We have to make sure this Majoron does not have a substantial
coupling to $Z^0$ boson in order to avoid the stringent constraints from
the LEP experiment.  In particular, we need to make sure the
real partner of Majoron is not too light.
In our scheme, $M_\nu$ arises at tree level through the VEV of a
$SU(2)$ triplet Higgs  $\Delta^{\!\!\scriptscriptstyle \alpha\beta}$, with
hypercharge $Y=1$ and lepton number $L=-2$.  To avoid the LEP
constraint, one can break lepton number using another Higgs boson with
much larger VEV so that the Majoron effectively does not couple to
$Z^0$ at tree level.  A nice choice is another $SU(2)$ triplet
$T^{\scriptscriptstyle \alpha}_{\scriptscriptstyle \beta}$
with hypercharge $Y=0$ and lepton number $L=2$.  We will also
discuss a less favorable case with $T$ being replaced by a singlet
$\sigma$, carrying the same hypercharge and lepton number.
The triplet ($T$), however, is a better choice because
the singlet ($\sigma$) will naturally couple to the singlet neutrinos
($\nu_{\scriptscriptstyle S}$'s) and give
rise to large $M$ in Eq.(1).   The resulting Higgs sector of the
model we consider here is close to those in Refs.\cite{MS,GR,CMP,sigma}.
Most of the result in  this section could be borrowed directly
from the references.  As we shall see below, the model naturally gives
rise to very small $M (\ll M_\nu)$.

The full scalar potential is given by
\def\tr{{\rm tr}\,}
\begin{eqnarray} \label{V}
V(\phi,\Delta,T)&=&\mu_2^2\phi^\dagger\phi+\mu_3^2\,
\tr(\Delta^\dagger\Delta)+
\lambda_1(\phi^\dagger\phi)^2+\lambda_2[\tr(\Delta^\dagger\Delta)]^2
\nonumber\\
&&+\lambda_3\phi^\dag\phi\,\tr(\Delta^\dag\Delta)+
\lambda_4\,\tr(\Delta^\dag\Delta\Delta^\dag\Delta)+
\lambda_5(\phi^\dag\Delta\Delta^\dag\phi)\nonumber\\
&&+\mu_1^2\,\tr(T^\dag T)+\zeta_1[\tr(T^\dag T)]^2+
\zeta_2(\phi^\dag\phi)\tr(T^\dag T)
+\zeta_3\,\tr(T^\dag T T^\dag T) \nonumber\\
&&+\zeta_4\,\tr(\phi^\dag T T^\dag  \phi) +
\zeta_5\,\tr(\Delta^\dag\Delta)\tr(T^\dag T) +
\zeta_6\,\tr(\Delta^\dag\Delta T^\dag T)  \nonumber\\
&& + \zeta_7\,\tr(\Delta^\dag T^\dag \Delta T)
- \kappa(\phi^{\scriptscriptstyle \alpha} \phi^{\scriptscriptstyle \beta}
  \Delta^\dag_{\scriptscriptstyle \alpha\gamma}
T^{\scriptscriptstyle \gamma}_{\scriptscriptstyle \beta}+{\rm h.c.}) \; ,
\end{eqnarray}
where explicit $SU(2)$ indices are shown only for the $\kappa$-term.
The singlet version has $T$ replaced by $\sigma$.
Note that the $\kappa$ coupling naturally give rise to the relation among
the VEV's as given by
\begin{equation} \label{kv3}
v_{\scriptscriptstyle 3} \sim \kappa v_{\scriptscriptstyle 2}^2
v_{\scriptscriptstyle T} / \mu_{\scriptscriptstyle 3}^2 \; ,
\end{equation}
where $v_i$ are defined as
\beqa
\frac{v_{\scriptscriptstyle T}}{\sqrt{2}}
&\equiv& \lla T \rra \quad ({\rm or} \;\;\lla \sigma \rra ) \; ; \nonumber \\
\frac{v_{\scriptscriptstyle 2}}{\sqrt{2}} &\equiv& \lla \phi \rra \; ; \nonumber \\
\frac{v_{\scriptscriptstyle 3}}{\sqrt{2}} &\equiv& \lla \Delta \rra \; .
\eeqa
The VEV's are understood to be real and lie in the directions of the neutral
components. The neutral scalar mass matrix is given by
\begin{equation} \label{scalar}
\mbox{\boldmath $M_{\!\scriptscriptstyle R}^2$}=\left(\begin{array}{ccc}
2(\zeta_1+\zeta_3)v_{\scriptscriptstyle T}^2
+\frac{1}{2}\kappa v_{\scriptscriptstyle 2}^2
\frac{v_{\scriptscriptstyle 3}}{v_{\scriptscriptstyle T}} &
(\zeta_2+\zeta_4) v_{\scriptscriptstyle T}v_{\scriptscriptstyle 2}
-\kappa v_{\scriptscriptstyle 2}v_{\scriptscriptstyle 3} &
(\zeta_5+\zeta_6+\zeta_7) v_{\scriptscriptstyle T}v_{\scriptscriptstyle 3}
-\frac{1}{2}\kappa v_{\scriptscriptstyle 2}^2 \\
(\zeta_2+\zeta_4) v_{\scriptscriptstyle T}v_{\scriptscriptstyle 2}
-\kappa v_{\scriptscriptstyle 2}v_{\scriptscriptstyle 3} &
2\lambda_1v_{\scriptscriptstyle 2}^2 &
(\lambda_3+\lambda_5) v_{\scriptscriptstyle 2}v_{\scriptscriptstyle 3}
-\kappa v_{\scriptscriptstyle T}v_{\scriptscriptstyle 2} \\
(\zeta_5+\zeta_6+\zeta_7) v_{\scriptscriptstyle T}v_{\scriptscriptstyle 3}
-\frac{1}{2}\kappa v_{\scriptscriptstyle 2}^2 &
(\lambda_3+\lambda_5)v_{\scriptscriptstyle 2}v_{\scriptscriptstyle 3}
-\kappa v_{\scriptscriptstyle T}v_{\scriptscriptstyle 2} &
2(\lambda_2+\lambda_4)v_{\scriptscriptstyle 3}^2
+\frac{1}{2}\kappa v_{\scriptscriptstyle 2}^2
\frac{v_{\scriptscriptstyle T}}{v_{\scriptscriptstyle 3}}
\end{array}  \right) \; ,
\end{equation}
The neutral pseudo-scalar mass matrix is given by
\begin{equation}
\mbox{\boldmath $M_{\!\scriptscriptstyle I}^2$}=\left(\begin{array}{ccc}
\frac{1}{2}\kappa v_{\scriptscriptstyle 2}^2
\frac{v_{\scriptscriptstyle 3}}{v_{\scriptscriptstyle T}} &
\kappa v_{\scriptscriptstyle 2}v_{\scriptscriptstyle 3} &
\frac{1}{2}\kappa v_{\scriptscriptstyle 2}^2 \\
\kappa v_{\scriptscriptstyle 2}v_{\scriptscriptstyle 3} &
2\kappa v_{\scriptscriptstyle T}v_{\scriptscriptstyle 3} &
\kappa v_{\scriptscriptstyle T}v_{\scriptscriptstyle 2} \\
\frac{1}{2}\kappa v_{\scriptscriptstyle 2}^2 &
\kappa v_{\scriptscriptstyle T}v_{\scriptscriptstyle 2} &
\frac{1}{2}\kappa v_{\scriptscriptstyle 2}^2
\frac{v_{\scriptscriptstyle T}}{v_{\scriptscriptstyle 3}}
\end{array}  \right)\;,
\end{equation}
which has two zero mass eigenvalues, as expected. The only non-zero mass
eigenvalue of {\boldmath $M_{\!\scriptscriptstyle I}^2$} is given by
\begin{equation}
m_{\scriptscriptstyle A}^2=\frac{1}{2}\kappa
\frac{v_{\scriptscriptstyle 2}^2
v_{\scriptscriptstyle T}^2+v_{\scriptscriptstyle 2}^2
v_{\scriptscriptstyle 3}^2+4v_{\scriptscriptstyle 3}^2
v_{\scriptscriptstyle T}^2}
{v_{\scriptscriptstyle 3}v_{\scriptscriptstyle T}} \; ,
\end{equation}
which would be heavy, at least around the EW scale.
The massless Majoron can be made massive by introducing
an explicit soft lepton number violating $\mu_{\scriptscriptstyle T}^2
(TT+T^\dag T^\dag ) $ term to the scalar potential in Eq.(\ref{V}), if
this is necessary to evade astrophysical or cosmological constraints.
Note that the term affects the vacuum solution of $V$ but not the form of
{\boldmath $M_{\!\scriptscriptstyle R}^2$}.

With a light majoron,  a small eigenvalue
from {\boldmath $M_{\!\scriptscriptstyle R}^2$} (corresponding to
a light physical scalar),  has to be avoided in order
not to change the invisible width of the $Z^0$-boson decay beyond the
stringent experimental bound. An alternative way of making all the scalars
heavy is to impose the hierarchy $v_{\scriptscriptstyle 3} \ll
v_{\scriptscriptstyle T}$. In the latter case, the majoron, as well as the
potentially light scalar, will be predominantly the $T^0$ or $\sigma$
state, which does not coupled to $Z^0$.
A careful inspection of Eq.(\ref{scalar}) shows the hierarchy,
$v_{\scriptscriptstyle 3} \ll v_{\scriptscriptstyle T}$, is necessary in
order to avoid a scalar of mass smaller
than $\sim \sqrt{v_{\scriptscriptstyle 3}v_{\scriptscriptstyle 2}}$.
Assuming the hierarchy, Eq.(\ref{scalar}) gives eigenvalue for
the predominantly $\Delta^0$ state of the order
$\kappa v_{\scriptscriptstyle 2}^2
\frac{v_{\scriptscriptstyle T}}{v_{\scriptscriptstyle 3}}$,
hence above EW scale. The other two eigenvalues are at least of order
$v_{\scriptscriptstyle T}^2$ and $v_{\scriptscriptstyle 2}^2$
respectively,
with the latter corresponding to the predominantly $\phi^0$ state.

{\it Neutrino masses and model parameters.} Take the
$\mu\,$---$\,\nu_\mu$ family parameters. The Dirac mass $D$ in
Eq.(\ref{mass}) has to be suppressed relative to charged lepton
mass by roughly an order of $\epsilon = m_{\nu_\mu} / m_{\mu} \sim
10^{-8}$. Lepton number violating Majorana entries to mass matrix
in (\ref{mass}) contribute to the diagonal blocks, $M_\nu$ and
$M$. Entries to $M_\nu$ come from couple of
$\nu_{\scriptscriptstyle L}$'s to the scalar $\Delta$. Here we
required ${v_{\scriptscriptstyle 3}} \ll 10 \;\mbox{eV}$, from a
${v_{\scriptscriptstyle T}}$ bound obtained below. In
Ref.\cite{MS}, it has been illustrated, for the case without the
extra $T$ or $\sigma$, that the scalar $\Delta$ could be naturally
heavy and yet with a small VEV; in that case, the crucial term in
the scalar potential is the $\kappa$ term in $V$ [Eq.(\ref{V})]
with $T$ being replaced by a heavy mass parameter. In our modified
case, the same story goes with the $T$-VEV,
${v_{\scriptscriptstyle T}}$, playing the latter role. It can
easily be shown that appropriate choice of $\mu_i$ value in $V$
can fix the required hierarchy ${v_{\scriptscriptstyle 3}} \ll
{v_{\scriptscriptstyle T}}$ without fine-tuning. In fact, one
needs basically only a distinct $\mu_{\scriptscriptstyle 3}$ scale
being larger than that of the other mass parameters in the scalar
potential, as illustrated by Eq.(\ref{kv3}) obtained from
minimizing the scalar potential. The extra scalar $T$, or
$\sigma$, is necessary here as the $\kappa$ term without the
latter explicitly violates lepton number. The lepton number
violating Majorana mass terms for ${\nu_{\scriptscriptstyle
S}}$'s, being of the same dimension, would then derive divergent
loop contributions ruining our neutrino mass scheme.

If the singlet scalar $\sigma$ is used, it could couple directly
to singlet neutrinos, the $\nu_{\scriptscriptstyle S}$'s, and give
rise to neutrino mass after it develops VEV. If this contribution
is not suppressed, it will give rise to potentially larger
contribution than those from $\lla \Delta \rra$. This is one of
the reasons that we consider the version with the $T$ triplet
instead more interesting. The implementation of lepton number
violation through the triplet $T$-VEV  ensures $M\ll M_\nu$ as any
contribution to $M$ has to go through loop diagrams where a
$T$-VEV still has to be incorporated.

The $T$ scalar is actually a
very interesting EW triplet. Its neutral component $T^0$ does not couple
to $Z^0$, helping to get around the usual constraint on a majoron from a
nontrivial $SU(2)$ multiplet; while its $W$-boson coupling, or contribution
to the precision $\rho$-parameter, restricts $\lla T \rra$ to be
$< 0.04 \, {v_{\scriptscriptstyle 2}} \sim 10 \,\mbox{GeV}$\cite{CK}.
Hence, it predicts interesting accessible phenomenology for the majoron
and its real partner.

We will discuss below a Dirac-seesaw mechanism for the natural suppression
of the neutrino Dirac mass scale below that of the charged leptons, without
introducing particularly small Yukawa couplings. Following the idea, one can
simply assume the Dirac mass generating Yukawa couplings among the
leptonic-doublets and singlet neutrinos to be about the same, for
each of the three families, as those of the corresponding charged leptons.
If we further take the $M_\nu$ generating Yukawa couplings,
those among the leptonic doublets and $\Delta$, to be about the same as that of
$\phi^\dag$ leptonic Yukawa couplings, we would then have
$\Delta m_{\nu_\mu}/m_\mu \sim v_{\scriptscriptstyle 3}/
v_{\scriptscriptstyle 2}$; hence $v_{\scriptscriptstyle 3} \sim 10^{-11}\;
v_{\scriptscriptstyle 2}$. If we identify only the third family Yukawa couplings
in the latter case, and assume the suppressions of the Yukawa couplings
involving the $\Delta$ down the families as going as the square of thse
involving $\phi^\dag$, we would have $v_{\scriptscriptstyle 3} \sim 10^{-11}\,
v_{\scriptscriptstyle 2}\; \frac{m_\tau}{m_\mu}$, about an order of
magnitude
larger. This latter case is in accordance with assumptions used in our illustrative
neutrino mass spectrum at the beginning. From Eq.(\ref{kv3}), we have then the
required scale for $\mu_{\scriptscriptstyle 3}$ to be about
$10^{7}\,\mbox{GeV}$. As for $v_{\scriptscriptstyle T}$, $\sim 10^{-2} \;
v_2$ could be a reasonable estimate.

{\it A Dirac-seesaw mechanism.}
We introduce here a Dirac-seesaw mechanism to achieve at the Dirac mass
suppression factor $\epsilon$. Consider the Yukawa coupling
$ \bar{\nu}_{\scriptscriptstyle S} \phi^\dag {\nu}_{\scriptscriptstyle L}$
to be forbidden by a $Z_2$ symmetry under which only
${\nu}_{\scriptscriptstyle S}$ transforms non-trivially. The Dirac mass
term can be recovered through a dimension five term
$S\, \bar{\nu}_{\scriptscriptstyle S} \phi^\dag {\nu}_{\scriptscriptstyle L}$
with a  VEV for the scalar $S$ also  transforming non-trivially
under the $Z_2$ symmetry. We have then  the suppression factor
$\epsilon = \frac{\lla S \rra}{M_{\scriptscriptstyle N}} \sim 10^{-8}$,
where
$M_{\scriptscriptstyle N}$ corresponds to some relevant higher mass
scale. For instance, if there is a vector-like pair of singlet fermions
$N_{\scriptscriptstyle L}-N_{\scriptscriptstyle R}$ transforming trivially
under the $Z_2$ symmetry with Dirac mass ${M_{\scriptscriptstyle N}}$.
$N_{\scriptscriptstyle R}$ would couple to ${\nu}_{\scriptscriptstyle L}$
through $ \phi^\dag$ while $N_{\scriptscriptstyle L}$ would couple to
${\nu}_{\scriptscriptstyle S}$ through $S$. The former could be taken to be
of the same order as that of the charged lepton masses, denoted by
$m_{\scriptscriptstyle \ell}$, with a similar heirarchy down the families.
The scheme then results in a Dirac-seesaw mass matrix of the form
\[ \left(\begin{array}{cc}
0 &  \lla S \rra \\
 m_{\scriptscriptstyle \ell}  & M_{\scriptscriptstyle N}
\end{array}\right)\; .  \]
Integrating out $N_{\scriptscriptstyle L}-N_{\scriptscriptstyle R}$
gives the effective Dirac mass entries, $D$,
with the suppression as required.  Note that
$N_{\scriptscriptstyle L}-N_{\scriptscriptstyle R}$
here carry natural lepton numbers.

The model requires the scales $v_{\scriptscriptstyle T}$,
$v_{\scriptscriptstyle 2}$, $\mu_{\scriptscriptstyle 3}$ (which controls the
size of the VEV $v_{\scriptscriptstyle 3}$), $\lla S \rra$ and
$M_{\scriptscriptstyle N}$.  For a realistic model,
one may wish to reduce the number of necessary scales.  The scales
$v_{\scriptscriptstyle T}$ and $v_{\scriptscriptstyle 2}$
are sufficiently close to each other that they can be consider
as one scale.  The two intermediate scales $\mu_{\scriptscriptstyle 3}$ and
$\lla S \rra$ can
also be identified with each other, with both of order $10^{7} GeV$.  In
that case, $M_{\scriptscriptstyle N}$ could be around GUT scale, $10^{15}
GeV$.  These are just typical scales that make realistic embedding of the
model into a grand unification theory possible.  We shall not try to
provide such embedding here.

The Dirac-seesaw mechanism will not mess up with the general scheme of
the model. Any Majorana mass has to arise from lepton number violating
VEV(s). Consider $\lla \Delta \rra$ and $\lla T \rra$. They do not have
direct couplings to the singlet fermions. We check that contributions to
Majorana mass of the latter do not arise till at least two-loop level.
Moreover,  it can easily be shown that Majorana mass terms
for the $N_{\scriptscriptstyle L}$ and $N_{\scriptscriptstyle R}$
singlets contribute to $M_\nu$ or $M$ only with extra suppression factors
$\frac{v_{\scriptscriptstyle 2}}{M_{\scriptscriptstyle N}}$
and $\frac{\lla S \rra\! ^2}{M_{\scriptscriptstyle N}^2}$
respectively, as a result of the seesaw structure.

Moreover, the $Z_2$ symmetry can easily be modified to forbid the
undesirable couplings giving rise to Majorana mass for the
$\nu_{\scriptscriptstyle S}$'s, for instance from $\lla \sigma \rra$.
Afterall, the $Z_2$ symmetry is introduced here only as an explicit
illustration of the Dirac-seesaw mechanism needed to suppressed the entries
to $D$. More complicated symmetries can be used for the purpose,
which may reduce the required ratio of
$\frac{\lla S \rra }{M_{\scriptscriptstyle N}}$, by making $\epsilon
\sim \left(\frac{\lla S \rra }{M_{\scriptscriptstyle N}}\right)^{\!n}$,
and may be even take care of the family hierarchy itself. There are
plenty of horizontal/family symmetry models of the type
in the literature\cite{unc}.

{\it Conclusion.} We analyzed here the scenario of three family of
pseudo-Dirac neutrinos. An explicit model is presented to show how the
idea could work with an extended Higgs sector and an approximate lepton
number symmetry, broken spontaneously or otherwise. The $T$ triplet
version of our model promised a safe triplet majoron with interesting
phenomenology. A Dirac-seesaw mechanism is introduced separately
for the suppression of the Dirac masses. Higgs/majoron
phenomenology and the incoporation of the Dirac seesaw into a complete
horizontal symmetry model are the interesting issues to be further pursued.

{\it Acknowledgements : }
O.K. thanks the National Center for Theoretical Sciences,
and National Tsinghua University of Taiwan
for hospitality. DC is supported by a grant from National Science
Council of R.O.C. and O.K. was supported in part by the U.S.
Department of Energy, under grant DE-FG02-91ER40685.

%\noalign{\hbox{and}}

%\clearpage


\bigskip
\bigskip


\begin{thebibliography}{99}

\bibitem{RRR}
P. Ramond, R.G. Roberts, and G.G. Ross,
Nucl. Phys. {\bf 406} (1993) 19,
 P.H. Frampton and O.C.W. Kong, Phys. Rev. {\bf D55} (1997) 5501.
\bibitem{BKS}
J.N. Bahcall, P.I. Krastev, and A. Yu. Smirnov,
Phys. Rev. {\bf D58} (1998) 096016;
M.C. Gonzalez-Garcia, P.C. de Holanda, C. Pe\~na-Garay,
and J.W.F. Valle, FTUV/99-41, hep-ph/9906469.
\bibitem{sK}
Super-Kamiokande Collaboration, Y.Fukuda {\it et al.},
 Phys. Rev. Lett. {\bf 81} (1998) 1562;
 P. Lipari, hep-ph/9904443;
G.L. Fogli, E. Lisi, A. Marrone, and G. Scioscia,
Phys. Rev. {\bf D59} (1999) 033001.
\bibitem{LSND}
LSND Collaboration: A. Athanassopoalos {\it et. al.},
Phys. Rev. {\bf C54} (1996) 2685; Phys. Rev. Lett. {\bf 77} (1996) 3082.
\bibitem{4nu}
For analysis of four neutrino scenario, see, for example,
V. Barger, T.J. Weiler, and K. Whisnant, Phys. Lett. {\bf B427} (1998) 97;
V. Barger, S. Pakvasa, T.J. Weiler, and K. Whisnant,
Phys. Rev. {\bf D58} (1998) 093016.
\bibitem{EIR}
See, for example,
J.K. Elwood, N. Irges, and P. Ramond, Phys. Rev. Lett. {\bf 81} (1998) 5064.
\bibitem{kong}
For a discussion of the issue, within the context of supersymmetry without
R-parity,  see O.C.W. Kong, Mod. Phys. Lett. A. {\bf 14} (1999) 903.
\bibitem{gut}
For an example under a GUT framework, see
 Z. Berezhiani and A. Rossi, JHEP {\bf 9903} (1999) 002.
\bibitem{3nuR}
For a non-GUT example, see R.N. Mohapatra, hep-ph/9903261.
\bibitem{BBN}
D.N. Schramm and M.S. Turner, Rev. Mod. Phys. {\bf 70} (1998) 303.
\bibitem{FV}
R. Foot and R.R. Volkas, Phys. Rev. Lett. {\bf 75} (1995) 4350.
\bibitem{FTV}
R. Foot, M.J. Thomson, and R.R. Volkas, Phys. Rev. {\bf D53} (1996)
5349.
\bibitem{xSF}
R. Foot and R.R. Volkas, Phys. Rev. {\bf D55} (1997)
5147; R. Foot, Astropart. Phys. {\bf 10} (1999), 253;
P. Di Bari, P. Lipari, and M. Lusignoli, hep-ph/9907548.
\bibitem{SF}
X. Shi and G.M. Fuller,  Phys. Rev. {\bf D59} (1999) 063006.
\bibitem{DHPS}
A.D. Dolgov, S.H. Hansen, S. Pastor, and D.V. Semikoz, TAC-1998-028,
hep-ph/9809598.
\bibitem{KS}
P.J. Kernan and S. Sarkar,  Phys. Rev. {\bf D54} (1996) 3681.
\bibitem{CFV}
R.M. Crocker, R. Foot, and R.R. Volkas,
UM-P-99/11, hep-ph/9905461.
\bibitem{pD}
M. Kobayashi, C.S. Lim, and M.M. Nojiri,
Phys. Rev. Lett. {\bf 67} (1991) 1685;
C. Giunti, C.W. Kim, and U.W. Kim,
Phys. Rev. {\bf D46} (1992) 3034;
A. Geiser, Phys. Lett. {\bf B444} (1998) 358.
\bibitem{mir}
R. Foot, H. Lew, and R.R. Volkas, Phys. Lett.
{\bf B272} (1991) 67; Mod. Phys. Lett.
{\bf A7} (1992) 2567; R. Foot, Mod. Phys. Lett.
{\bf A9} (1994) 169;  R. Foot R.R. Volkas, Mod. Phys. Rev.
{\bf D52} (1995) 6595.
\bibitem{pDm}
J. Bowes and R.R. Volkas, J. Phys. {\bf G24} (1998) 1249;
P. Langacker, Phys. Rev. {\bf D58} (1998) 093017;
Y. Koide and H. Fusaoka, Phys. Rev. {\bf D59} (1999) 053004;
Z. Chacko and R. Mohapatra, hep-ph/9905388.
\bibitem{MS}
E. Ma and U. Sarkar,  Phys. Rev. Lett. {\bf 80} (1998) 5716.
\bibitem{afs}
A. Antaramian, L.J. Hall, and A. Ra\v{s}in,
Phys. Rev. Lett. {\bf 69} (1992) 187; see also
T.P. Cheng and M. Sher, Phys. Rev. {\bf D35} (1987) 3484.
\bibitem{GR}
G. B. Gelmini and M. Roncadelli, Phys. Lett. {\bf B99} (1981) 411.
\bibitem{CMP}
Y. Chikashige, R. Mohapatra, and R. Peccei,
Phys. Lett. {\bf 98B} (1980)  265.
\bibitem{sigma}
J. Schechter and J. W. F. Valle, Phys. Rev. {\bf D25} (1982) 774,
A.S. Joshipura, Int. J. Mod. Phys. {\bf A7} (1992) 2021,
M.A. D\'iaz, M.A. Garc\'ia-Jare\~no, D.A. Restrepo, and J.W.F. Valle,
Nucl. Phys. {\bf B527} (1998) 44.
\bibitem{CK}
D. Chang, W.Y. Keung, and P.B. Pal,
 Phys. Rev. Lett. {\bf 61} (1988) 2420;
D. Chang and W.Y. Keung,
 Phys. Rev. {\bf 39} (1989) 1386.
\bibitem{unc}
See for example, O.C.W. Kong, {\it\ Ph.D. dissertation --- UNC-CH}
(1997), {\it UMI-98-18361-mc}, and references therein.

\end{thebibliography}
\end{document}